# D2D Communications in LoRaWAN Low Power Wide Area Network: From Idea to Empirical Validation

Konstantin Mikhaylov, Juha Petäjäjärvi, Jussi Haapola and Ari Pouttu
Centre for Wireless Communications
University of Oulu, P.O. BOX 4500, Oulu, Finland
Email: firstname.lastname@oulu.fi

*Abstract*—In this paper we advocate the use of device-to-device (D2D) communications in a LoRaWAN Low Power Wide Area Network (LPWAN). After overviewing the critical features of the LoRaWAN technology, we discuss the pros and cons of enabling the D2D communications for it. Subsequently we propose a network-assisted D2D communications protocol and show its feasibility by implementing it on top of a LoRaWAN-certified commercial transceiver. The conducted experiments show the performance of the proposed D2D communications protocol and enable us to assess its performance. More precisely, we show that the D2D communications can reduce the time and energy for data transfer by 6 to 20 times compared to conventional LoRaWAN data transfer mechanisms. In addition, the use of D2D communications may have a positive effect on the network by enabling spatial re-use of the frequency resources. The proposed LoRaWAN D2D communications can be used for a wide variety of applications requiring high coverage, e.g. use cases in distributed smart grid deployments for management and trading.

## I. INTRODUCTION

The Low Power Wide Area Networks (LPWANs) represent an exciting new trend in the development of the wireless communication systems. In contrast with the traditional mobile broadband, these technologies have been developed with focus on machines and their requirements. Machine-type communications have governed the very design of these technologies and their key performance indicators. Namely, the common factors for most LPWAN technologies of today are 1) support for massive deployment of low traffic end devices (EDs); 2) low cost of EDs; and 3) low cost of infrastructure in production and exploitation. The general approach for addressing the above factors by the today's LPWAN technologies is to employ multi-kilometer wireless communication links to achieve good scalability and to minimize the infrastructure investments, as well as to maximally simplify devices and protocols to reduce cost and energy expenditure of EDs. The approach enables the use of LPWAN technologies for broad range of possible applications, e.g. nature or infrastructure monitoring, assets or people tracking, and, importantly, smart energy metering and smart grids monitoring.

The landscape of today for the LPWAN technologies is excessively diverse and consists of numerous proprietary solutions, including Sigfox, Ingenu/On-Ramp, Starfish, Cynet, Accellus, Telensa, etc. Furthermore, several open and semi-open technologies (e.g., LoRaWAN, Weightless), and few standardized technologies (IEEE 802.15.4k or LTE-M and NB-IoT) exist. Unfortunately, information about many of these technologies, especially the ones belonging to the group of proprietary technologies, is extremely scarce. Scarcity of information substantially hampers the research activities making the studies primarily focus on the few technologies for which information can be obtained. One such technology is the Long Range (LoRa) Wide Area Network (WAN).

A few comparisons of the LoRaWAN with the other contemporary LPWAN technologies is provided in [1]-[3]. The results of experimental studies of the communications performance and coverage have been presented, e.g. in [4]-[6]. The performance of LoRa in highly dynamic environment and the effect of the various spreading factors (SF) on the range of indoor communication together with the energy consumption of a LoRaWAN devices were empirically studied and reported in [7] and [8], respectively. With respect to the network behavior of the technology, the fundamental performance limitations of the technology (e.g., maximum throughput) as well as the maximum number of the devices serviceable by a single LoRaWAN gateway for few characteristic applications were analyzed in [9]. In [10] tools of stochastic geometry were employed to characterize the performance of a LoRa network subject to inter-network interferences. The authors of [10] show that the performance decays exponentially with the increase of number of devices in the network. The authors of [11] address scalability of a LoRa network showing, by means of simulations, that the interferences may have very strong effect on the network performance. They also show that densification of the gateways can help mitigating this problem, and that the approach is more efficient than use of directional antennas at the gateways. Empirical results highlighting the effect of the interference on LoRa-based communications are reported in [12] and [13]. The authors of [12] experimentally show the presence of the capture effect, which enables reception of the packets under interfering signal with the same SF. These results are used to assess the number of devices which can be managed by a single and by multiple gateways. In [13] results of an extensive experimental study on effect of the in-network interferences on LoRa network's performance are reported. The effect of the interference between all eight modulation coding schemes (MCS, referred to as data rates (DRs) in LoRaWAN) allowed in the EU 868 MHz band on the communication performance is investigated.

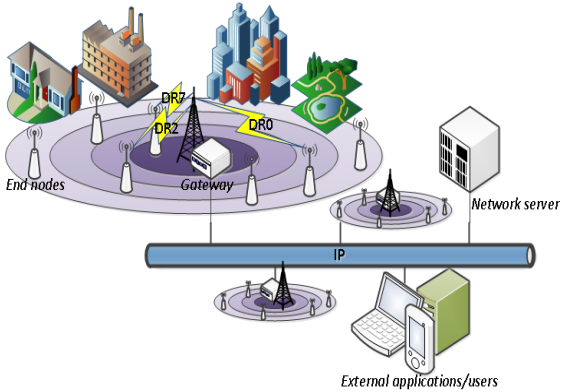

Fig. 1. Typical topology of a LoRaWAN network.

A few critical observations can be derived from the pointed studies. The possibility of a single LoRaWAN to operate sufficiently well in the broad range of conditions is proven by selecting one of the available DRs, featuring different tradeoffs between the on-air time and the communication range. Further, the scalability of a LoRaWAN network is somewhat limited due to the use of Aloha-based channel access protocol, which arises from the need to obey rather strict duty-cycle restrictions imposed by frequency regulation authorities. Duty-cycle limitations together with the requirement for having up to two receive windows after each uplink frame also drastically limit the technology with respect to the maximum achievable throughput for both uplink and, especially, for downlink. There are many applications for which these drawbacks may not appear to be crucial, however there are others, e.g. ones implementing a full control loop based on uplink-downlink communications where the limitations may result in catastrophic performance degradation, especially once the network will start scaling up. Bearing the previous in mind, this paper advocates the possibility of implementing a network assisted device-to-device (D2D) communication protocol in LoRaWAN environment. The main contribution of this paper is the empirical proof of the feasibility of a network assisted D2D communications implementation in a LoRaWAN network, which can be implemented even with the currently commercially available devices. Moreover, we detail the mechanisms developed for enabling D2D communications and the results of real-life measurements detailing the energy and time benefits of D2D communications in comparison to the default communications mechanisms. To the best of authors' knowledge, this is the first paper reporting the D2D solution for the LoRaWAN network.

The rest of the paper is organized as follows. In Section II we briefly overview the main features of the LoRaWAN technology. In Section III we first advocate the D2D communications for LoRaWAN and discuss the pros and cons of the approach. Then we detail our proposed solution and its implementation, and report the results of the conducted experimental measurements. Finally, Section IV concludes the paper and highlights the major observations.

TABLE I. PARAMETERS OF LORAWAN DRS FOR EU 868 MHZ BAND [14]

| DR | Modu-lation | LoRa SF | Bandwidth, kHz | PHY rate, bps | Max MAC payload, bytes |
|---|---|---|---|---|---|
| 0 | LoRa | 12 | 125 | 250 | 59 |
| 1 | LoRa | 11 | 125 | 440 | 59 |
| 2 | LoRa | 10 | 125 | 980 | 59 |
| 3 | LoRa | 9 | 125 | 1760 | 123 |
| 4 | LoRa | 8 | 125 | 3125 | 230 |
| 5 | LoRa | 7 | 125 | 5470 | 230 |
| 6 | LoRa | 7 | 250 | 11000 | 230 |
| 7 | GFSK | 50 000 bit/s rate | | 50000 | 230 |

## II. LoRaWAN Technology

Coming to the market just a few years ago, today LoRaWAN has become one of the most visible LPWAN technologies. The technical solutions underlying the technology consist of two components: the network protocol and the LoRa modulation.

The LoRaWAN network protocol is defined in the LoRaWAN specification [15]. A typical LoRaWAN network is illustrated in Fig. 1. The network is built with a star-of-stars topology, where the EDs communicate with gateways (GW), which relay the data between EDs and the central network server (NS) connected through a backhaul IP-based connection. The EDs communicate with the GW wirelessly using one of the available DRs (the eight options for EU 868 MHz band are listed in Table I). The DRs feature various tradeoffs between the communications range and on-air time. The optional adaptive DR (ADR) feature can be used to enable NS controlling the power and the DR used by a particular ED. This may enable more efficient management of the radio resources on one hand, and bring forth energy savings on the other hand.

There are three types of the EDs currently defined for the LoRaWAN. Class A devices are entitled to send their data at any moment of time, thus implementing Aloha-type channel access, over one of the frequency channels. The used channel is selected randomly from the list of the channels supported by the GWs. Two receive windows (RWs) with the specific timings are opened after each uplink message for downlink communications. In addition to this, devices of class B upkeep synchronization with the GW and open RWs at specified time slots. Devices of class C do not employ low power modes and keep receiving whenever not transmitting. The implementation of class A functionality is obligatory for all LoRaWAN EDs, while class B and C functionalities are treated as optional. Notably, since the LoRaWAN devices do not employ any listen before talk technique, in many regions and bands (e.g. EU or China) they are subject to the duty-cycle based channel access restrictions imposed by corresponding frequency regulations. As a result, the EDs are required to track their own on-air time for each particular sub-band and back off transmission on this channel to comply with the regulations.

The LoRa modulation scheme developed by Semtech enables long-range communications. The modulation is based on a chirp spread spectrum scheme with chirp signal constantly changing the frequency [1]. The raw physical rate is thus defined as [1]:

$$R_b = SF \cdot BW/2^{SF},$$

where *BW* is the bandwidth (refer to Table I). To increase the probability of successful data transfer, the LoRa additionally employs error correcting codes (most often with the fixed rate of 4/5 [14]). The range of allowed transmit power in EU 868 MHz band ranges from 2 to 20 dBm.

Note, that the LoRaWAN technology has already been employed for energy grids and smart city management purposes. For example, plans for installing 400 000 smart meters operating with LoRaWAN have been announced in Russia [16]. A street light control solution based on this technology has been reported in [17].

## III. DEVICE-TO-DEVICE COMMUNICATIONS IN LORAWAN

### A. D2D in LoRaWAN: Feasibility, Pros, and Cons

Two principal questions arise for enabling D2D communications in LoRaWAN: 1) "Is there a need for it?"; 2) "Is this possible at all?". Below we address both of them.

There are several reasons to enable the D2D communications in LoRaWAN. The first one is the need for addressing cases where two neighboring devices have to exchange data (e.g. one can consider a sensor and an actuator implementing a control loop or an energy prosumer and consumer in a microgrid as discussed e.g., in [18]). In the current architecture (see Fig. 1), the above cases have to be implemented using the NS (probably involving a special app, which would read the sensory data from the NS and instruct the server to transfer this data back to the actuator) as mediator. In this respect, D2D-based communications in the context of LoRaWAN has three major advantages. 1) The possibility of using a higher DR for D2D communications between devices in close proximity, resulting in lower on-air time and lower energy consumption; 2) Lower latency and higher throughput of the D2D link compared with communications through a gateway due to synchronization of the EDs, as well as possibility of omitting the RWs (and respective delays) and communications over backhaul. This may enable new applications with more severe requirements, not feasible currently for LoRaWAN; 3) Reduction of the network and backhaul load is enabled. More precisely, the DR used in a D2D link between EDs of close proximity located at sufficient distance from the GW can be set to prevent any distortion for the rest of the network, thus enabling more efficient use of the radio spectrum.

The drawbacks of enabling the D2D communications can be listed as follows. First is the possible negative effect (interference) on the operation of the LPWAN (e.g., if the devices operate close to the GW or if they do not synchronize their operation with the GW). Second, the data sent over a D2D link are not registered by the NS in any way. Finally, these are various security and privacy concerns that are not covered here.

With respect to the feasibility of D2D communications implementation, we have to note that the most recent LoRaWAN specification (v1.0.2) does not feature a mechanism enabling the EDs to communicate directly. This forces us to state that no ready-made solution enabling LoRaWAN-compatible D2D communications exists. In addition, the LoRaWAN operates in the license-free industrial, scientific and medical (ISM) sub-GHz band, which can be utilized by any other device subject to obeying the respective regulations. Also we note that many of the current commercial LoRa/LoRaWAN chipsets (e.g., RN2483 or SX1272) feature some capabilities for low level control, which can provide the basis for D2D communications implementation.

Another principal question, in our opinion, is how to set-up a D2D communications link? The first option can be named the *ED-constructed D2D* and would imply that two EDs use a mechanism to detect each other (e.g., periodic advertisements) and establish the D2D link on their own. This approach has two substantial drawbacks. The first one is the overhead traffic and energy consumption. The second one is the potential interference caused to the in-network communication (given that the advertising and D2D link are done on the same channels and using non-orthogonal signals). The second approach, which appears to us more attractive, can be named *network assisted D2D*. This approach presumes that the D2D link construction should be done under the supervision of the NS, letting it (or a special application) to decide a) the possibility and feasibility of establishing a D2D link and b) the parameters (e.g., channel, power, time frame, DR or other MCS parameters) for the D2D link. Note that the initiative to form a D2D link may come either from the EDs or from the NS. The major advantages of this approach are a) low overheads (commands can even be merged with the conventional uplink/downlink traffic) and b) possibility of avoiding any negative impact on the network operation. Note, that as this is discussed, e.g., in [19], the LoRaWAN GWs have means to enable the localization of the EDs. This empowers the NS with the means to effectively assess the feasibility of a D2D link and select the parameters for it.

### B. Proposed LoRaWAN-D2D Protocol

In order to prove the feasibility and assess the possible performance of a D2D link in LoRaWAN, we opted for a simple protocol, illustrated in Fig. 2.

The D2D link is established by two class A EDs based on a command from the NS. To initiate the D2D establishment process, the NS transmits messages to the EDs to set up the link. Each of the messages include 1) the frequency (or a set of frequency channels and the data to determine the timing and hopping sequence), 2) the MCS parameters to be used, 3) the timing parameters for the start and duration of the D2D session, and 4) the information about the role of each device (to determine which of the devices has to send the first packet). Additionally, the message may include the addresses the EDs have to use, as well as the security keys (note that the LoRaWAN communications are encrypted). One substantial problem at this stage is how to synchronize the EDs, given that the devices of class A send their uplink packets at random and the downlink messages are sent in the RWs opened some time after an uplink packet. Furthermore, one cannot be certain that all the uplink/downlink packets are delivered. To address the above issues, the NS has to enable a substantial time gap between transmitting the messages and the start of the D2D event (i.e., the moment when initiator sends its packet). Also the NS should set the listening window long enough to maximize the probability of the rendezvous between the initiator (ED sending the first packet) and the scanner(s) (ED(s) waiting for the packet from the initiator). The optimal selection of these parameters, mindful of energy consumption and depending on the network situation and the operation settings of the EDs, is not a trivial problem, but for the purpose of this paper is considered out of scope and left for further study.

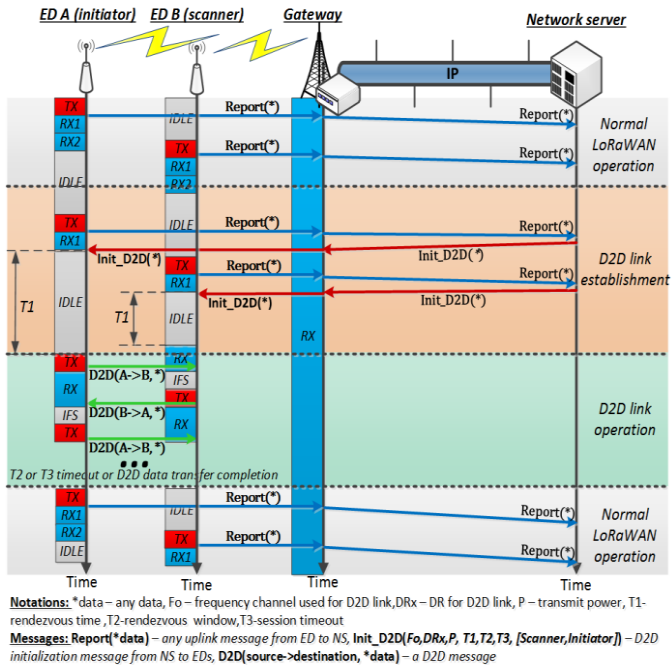

Fig. 2. Sequence of messages in the proposed network assisted D2D protocol (two EDs and one GW case, D2D in a single channel with constant DR).

After receiving these messages, the EDs reconfigure their transceivers with the parameters provided by the NS and prepare for a D2D session. At the specified time the scanners enable their receivers. In its turn, the initiator prepares the first packet and starts transmitting it at the prescribed time. If the scanner receives the packet from the initiator, the devices start alternating transmitting packets to each other. As can be seen, at the D2D phase, the proposed protocol is quite similar to Bluetooth. The link is closed either once all the data have been transferred (or once the link is not needed any more), or due to the no-reply or session duration timeouts. A timeout is also used to stop the idle listening by scanners and initiators in case the rendezvous has not occurred. After timeout the D2D link is considered to be cancelled and the devices reconnect to the NS.

Note that the described D2D phase protocol is just one possible option for a D2D link. A more advanced solution, perhaps employing some form of listen before talk (LBT) not restricted by duty cycle regulations may be considered but is currently not feasible by the used LoRaWAN hardware.

*C. Implementation and Evaluation*

To prove the feasibility of D2D in LoRaWAN in general and of the proposed solution in particular, a demonstrator has been instrumented. The demonstrator was built using the modular Wireless Sensor and Actuator Network (WSAN)/IoT platform developed at the Centre for Wireless Communications (CWC). To enable the LoRaWAN connectivity several shields hosting the RN2483 LoRaWAN transceivers[1] from Microchip (firmware version 1.0.1) were designed and produced. The two class A EDs composed of the main board with an ARM-based controller, a LoRaWAN module and, optionally, a battery board, were used in the test. The LoRaWAN connectivity was enabled by a LoRa Wirnet 868 MHz GW from Kerlink, which was configured to cooperate with the NS from Semtech[2]. The commands to initiate the D2D link were given to the NS from an internet-connected computer using the web-based interface. The distances from the EDs to the GW and between the two EDs were 2.1 km and 10 meters, respectively. The equipment used in the tests (except the GW) is depicted in Fig. 3.

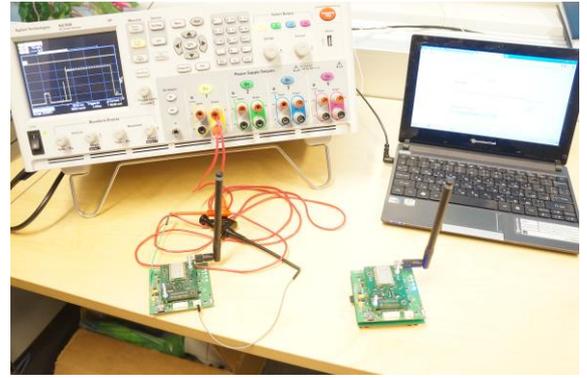

Fig. 3. Equipment used in tests: 2 EDs based on CWC's platform, a computer with web interface to Semtech NS, and a power analyser used to capture the consumption of an ED.

The EDs were programmed with the in-house developed firmware, composed of the FreeRTOS middleware and proprietary low-level interface drivers (refer to [20] and [23]), in-house developed RN2483 drivers supporting both operation of devices in LoRaWAN and the direct control over the radio, and the threads implementing the described D2D protocol and switching between LoRaWAN and D2D.

After startup procedures the EDs join the LoRaWAN network (using over-the-air activation procedure) and start sending their data to the NS with a period of about 5 seconds using DR0 (ADR disabled) and 14 dBm power. The commands to initiate the D2D link are given using the downlink message form of the Semtech NS's web interface. The following parameters for the D2D link were used: port=0xDD, Fo=865 MHz, DR=6, P=14 dBm, T1=0 for scanner and 15 s for initiator T2=30 s. Additionally, a 50 ms delay was introduced for the D2D communication between the reception of a packet and transmission of a response. To test the D2D functionality, the initiator transferred ten packets with 240-bytes of random payload each, and the scanner acknowledged each received packet with a short (10 byte) packet. Once the transmission was successfully completed (as well as in case of any D2D link failure) both EDs switched back to LoRaWAN.

*D. Discussion of the results*

The initial conducted tests have proved the feasibility of the proposed D2D mechanism. In more than 96.9% of the conducted experiments, the two EDs successfully established a D2D link. All the ten packets were delivered and the D2D was completed

---

[1] according to [22], RN2483 was the world's first transceiver to pass the LoRaWAN™ Certification Program

[2] http://iot.semtech.com/

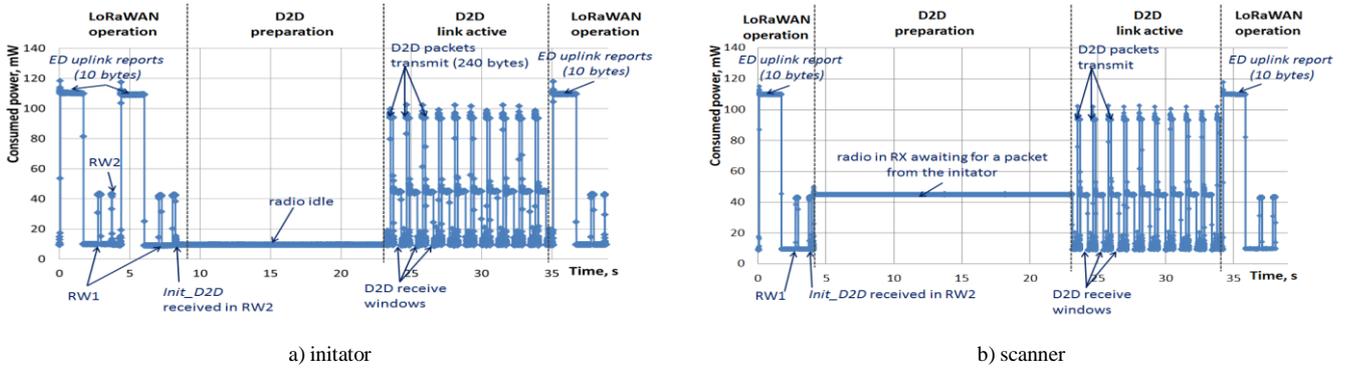

Fig. 4. Radio power consumption profiles during normal LoRaWAN operation and D2D link establishment and D2D session (note that the consumption of the main controller and the other peripherals has been filtered out).

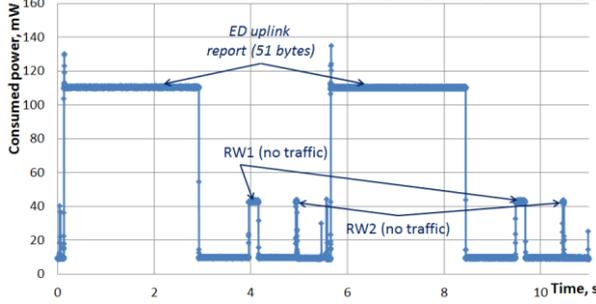
a) transmitter (51 byte packet, DR0, no downlink traffic)

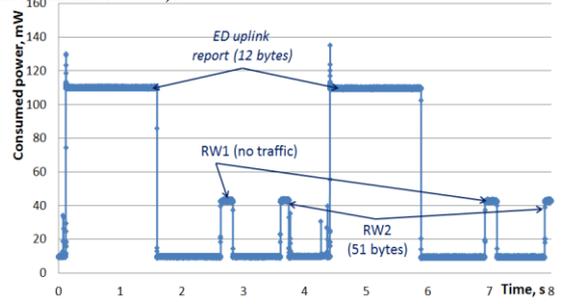
b) receiver (12 byte uplink packet at DR0, 51 bytes in downlink in RW2)

Fig. 5. Radio power consumption profile during LoRaWAN operation (consumption of the main controller excluded).

TABLE II. COMPARISON OF TIME AND ENERGY FOR TRANSFERRING 2400 BYTES OF DATA OVER CONVENTIONAL LoRaWAN AND THE PROPOSED D2D SCHEME

| Parameter | Conventional LoRaWAN operation (DR0) | | Proposed D2D mechanism (DR0 during initiation, DR6 for D2D) | |
|---|---|---|---|---|
| | Transmitter | Receiver | Initiator | Scanner |
| Time, s | 225.6[1] | | 30.2[2] | |
| Energy consumption, J | 15.696 | 9.120 | 0.817 | 1.494 |

[1] $T=floor(2400/51)*max(transmitter\ period, receiver\ period)$, note that the possible duty cycle restrictions are not accounted for
[2] D2D session time for the initiator, connection establishment phase included

successfully in 91% of the experiments. Also we confirmed that after a D2D session the LoRaWAN transceivers are able to continue their operation in LoRaWAN without any need to re-join the network (for the duration of D2D session the operation of LoRaWAN stack on RN2483 was suspended using *mac pause* and later re-enabled with *mac resume* command).

Once the basic functionality was proved, we proceeded with assessing the performance of the instrumented D2D solution with respect to its energy consumption and robustness. For this we have connected one of the nodes to a N6705 power analyzer from Agilent. The power consumption logs (3V supply voltage) captured for the initiator and the sender devices with some clarifications are presented in Figs. 4 a) and b), respectively. The similar results for conventional LoRaWAN operations are depicted in Fig. 5. Based on these results we have estimated the total time and energy required to transfer about 2400 bytes of data between the two EDs via a D2D link and by conventional LoRaWAN mechanisms. The respective results are summarized in Table II. Note that the results for the scanner device are subject to substantial deviation due to the variation of the idle listening duration. The results presented in Table II were obtained for the case of approximately 19 seconds listening prior to D2D session.

As one can see from the presented results, even with no optimizations for the protocol and parameters, the D2D transfer required 7 times less time and caused consumption of 6-20 times less energy. In case of higher data volumes this disproportion will grow even higher. Also note that in the current test we have not accounted for the effects of the duty cycle limitations, as well as the potential benefit to the LPWAN as a whole due to offloading the traffic from the network.

We consider these results to be rather complimentary, given that the protocol implemented by us is far from being optimal in any sense and can be improved in many ways. First, as this is discussed e.g., in [24], much more energy efficient and robust media access protocols exist and can be used during D2D establishment phase. Second, enablement of dynamic modification of D2D link parameters (e.g., DR or transmit power) as well as use of LBT-based solutions and advanced retransmission mechanisms may improve robustness and reliability of a D2D link. Finally, the EDs in our protocol stop communication with the NS for the whole duration of the D2D session. This may be undesirable for some use cases.

Finally we would like to highlight few practicalities, which were learned in the process of implementing and evaluating our D2D solution. First, one can clearly see that all the downlink packets are delivered by the GW during RW2 (configured with DR3 since lower DRs caused instable behavior of EDs while joining the network). Second, substantial problems were caused by the need of controlling the RN2483 transceivers using ASCII-based commands over a low-speed (57.6 kbps, use of rates over 100 kbps caused stability issues) UART interface.

This introduced very substantial overheads (energy and time wise) for communication between the controller and the transceiver, and hindered all attempts to enable any form of synchronization. Third, the used chipsets do not have any mechanism for measuring the power in a radio channel thus impeding implementation of any form of LBT. Finally, we witnessed that with ADR feature enabled the return to LoRaWAN operation after a D2D session was often not accomplished properly. Nonetheless, all these features are specific to the hardware used and do not invalidate the concept.

## IV. CONCLUSIONS

To the best of authors' knowledge the current paper reports the first attempt of proposing and implementing in practice the D2D communication in the context of LoRaWAN LPWAN technology. In the paper we first discussed some pros and cons of enabling D2D in LoRaWAN. We noted that enablement of D2D may have two significant positive effects. The former one is the reduction of the time and energy for data transfers between the involved devices. This may even become an enabler for new applications like, e.g., the distributed control in the smart energy grids. The other one is the release of substantial network resources. This may be especially beneficial in the case of massive device deployments, like the ones aimed at energy grids control. Next, we discussed the possibilities of implementing the D2D in LoRaWAN and proposed a network assisted D2D protocol. The proposed protocol has been implemented for the LoRaWAN certified commercial transceivers. The results of the experiments have confirmed the feasibility of the proposed solutions and enabled to assess its performance. Namely, we have shown that use of D2D link enables to reduce the time and energy for transferring 2400 bytes of data by 7, and 6-20 times, respectively. This would be beneficial in any control solution requiring low latency communications such as grid control in smart grids or industry automation to name a few. As the next step, we plan to take the proposed solution "in the field" and evaluate its feasibility in a real-life smart grid application.

The authors have to note that by no means the current paper gives the conclusive solution for implementing the D2D communication in LoRaWAN. In this paper we just show the feasibility of this and point out some of the possible solutions, as well as few open challenges. Among them is the problem of coming up with a theoretical framework enabling intelligent selection of the D2D link parameters, as well as of the links to be offloaded to D2D. Another more practical problem is the way to integrate the D2D with the current LoRaWAN architecture. The first question to address here is: "Whether the D2D should be managed by the NS, a special application, or in another way?" Another crucial challenge is development and introduction of the proper security and authentication mechanisms.

## V. ACKNOWLEDGEMENT

This paper describes work undertaken in the context of the P2P-SMARTEST project P2P-SMARTEST project, Peer to Peer Smart Energy Distribution Networks (http://www.p2psmartest-h2020.eu/), an Innovation Action funded by the H2020 Programme, contract number 646469.